\documentclass[aip,reprint,groupedaddress]{revtex4-1}
\pdfoutput=1

\usepackage{amsmath}
\usepackage[per-mode=symbol-or-fraction,
separate-uncertainty=true,multi-part-units=single]{siunitx}
\usepackage{graphicx}
\graphicspath{{figures/}}

\begin{document}

\title{Stimulus-responsive colloidal sensors with fast holographic readout}

\author{Chen Wang}
\author{Henrique W. Moyses}
\author{David G. Grier}
\affiliation{Department of Physics and Center for Soft Matter
  Research, New York University, New York, NY 10003}

\begin{abstract}
Colloidal spheres synthesized from polymer gels
swell by absorbing molecules from solution.
The resulting change in size can be monitored with nanometer
precision using holographic video microscopy.
When the absorbate is chemically similar to the polymer matrix,
swelling is driven primarily by the entropy of mixing, and
is limited by the surface tension of the swelling sphere
and by the elastic energy of the polymer matrix.
We demonstrate though a combination of optical micromanipulation and holographic
particle characterization that the degree of swelling of a single
polymer bead can be used to measure the monomer concentration 
\emph{in  situ}
with spatial resolution comparable to the size of the sphere.
\end{abstract}

\maketitle

Stimulus-responsive colloidal particles\cite{fujii05,*stuart10} respond to
physical or chemical changes in their environment through measurable
changes in their own physical properties.
Such particles have proved useful in 
a wide range of applications, ranging from drug-delivery systems
to probe particles for sensors.
Monitoring probe particles' responses can be challenging, particularly
for local probes involving changes in isolated particles.
Here, we demonstrate that in-line holographic microscopy can be used to
gauge the swelling of individual micrometer-scale
polymer-gel spheres \emph{in situ} and thus
to measure the local concentration of selected chemical species,
with excellent spatial and temporal resolution.
The key to this technique is the ability of
quantitative holographic video microscopy\cite{lee07a} 
to report the probe sphere's radius with nanometer
precision while simultaneously monitoring its refractive index
with part-per-thousand resolution.

To demonstrate the holographic concentration probe,
we combine holographic
micromanipulation\cite{dufresne98,*grier03}
with holographic video microscopy\cite{lee07,lee07a} to measure
concentration profiles of solubilized silicone oil in water.
Our probe particles consist of
polydimethylsiloxane (PDMS) synthesized by base-catalyzed hydrolysis
and copolymerization of difunctional dimethyldiethoxysilane (DMDES)
and trifunctional methyltriethoxysilane (MTES)\cite{obey94,*goller97}.
Synthesis and characterization of these particles is described
elsewhere\cite{wang15}.  When dispersed in pure water, these spheres
have a nominal radius of $a_0 = \SI{1}{\um}$, as measured by scanning
electron microscopy and 
\emph{in situ} holographic characterization\cite{lee07a,wang15}.
Trifunctional groups act as crosslinkers
for the PDMS gel, and the particles used in this study have
crosslinker fractions of $\xi = \num{0}$, \num{0.4}, and \num{0.8}.

PDMS gels absorb silicone oil and thus swell in the presence of
monomeric DMDES to a degree that depends
on the monomers' concentration in solution.
Measuring the sphere's radius through holographic microscopy
then provides a means to monitor the concentration of dissolved
DMDES in real time.

\begin{figure}[!t]
  \centering
  \includegraphics[width=0.8\columnwidth]{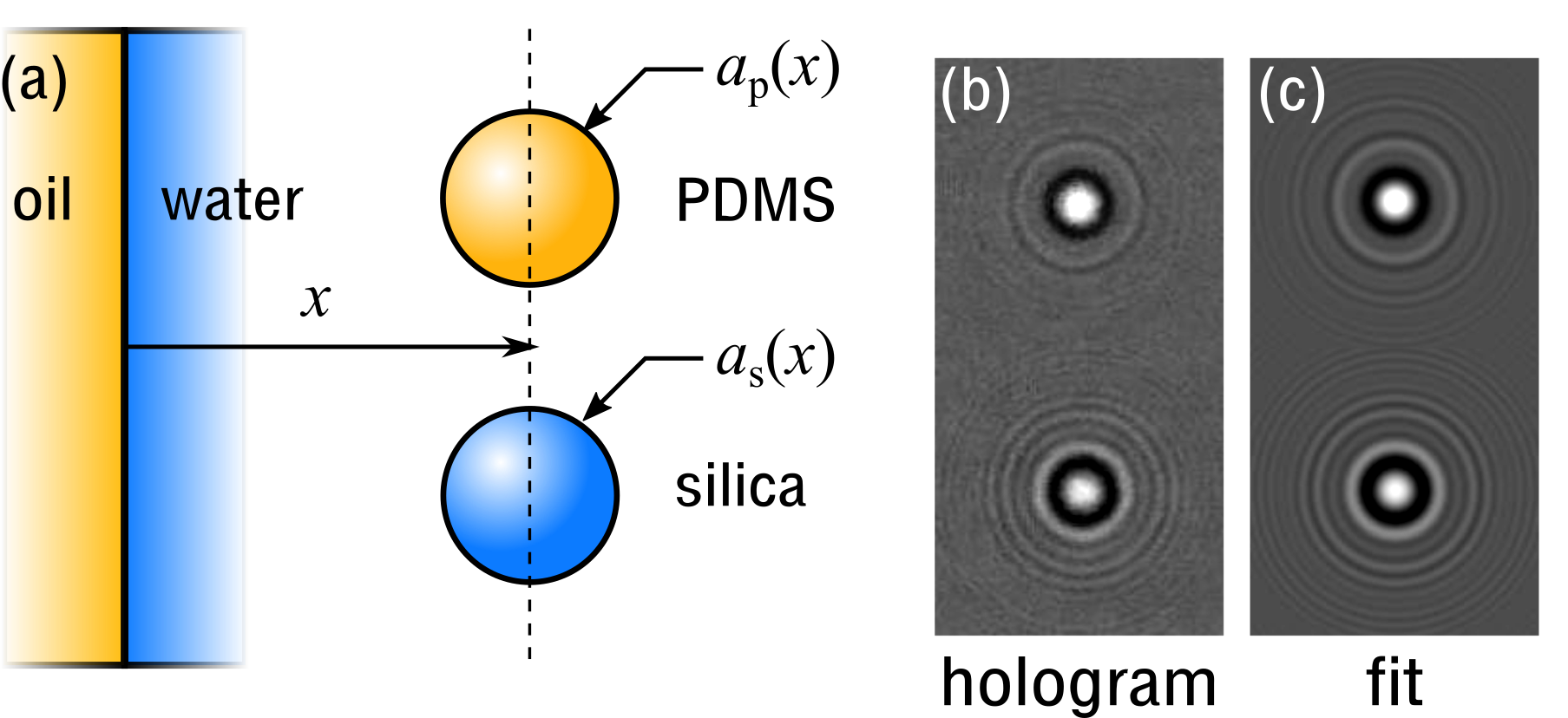}
  \caption{(a) Schematic representation of the local concentration
    measurement.  A polymer-gel (PDMS) sphere and a silica sphere
   are held by holographic optical tweezers at distance $x$ from an
   oil-water interface.  The spheres' radii and refractive indexes are
   measured by holographic video microscopy as a function of $x$.
   The PDMS sphere's properties probe the concentration
   profile, $c(x)$, of monomers in solution.  The silica sphere
   serves as a control.  (b) Measured
   hologram of the two spheres in solution.  (c) Fit to Lorenz-Mie
   theory for the positions and properties of the spheres.}
  \label{fig:schematic}
\end{figure}

Our system, depicted schematically in Fig.~\ref{fig:schematic},
consists of a mixture of PDMS spheres 
and silica spheres (Bangs Laboratories, Catalog number SS04N)
dispersed in a \SI{0.1}{M} solution of aqueous
ammonium hydroxide (Fisher Scientific).
This solution fills half the length of a \SI{2}{\cm}-long rectangular capillary
tube with $\SI{50}{\um} \times \SI{500}{\um}$ internal cross-section
(Vitrocom 5005).
The other half of the channel is filled with DMDES,
and a nearly planar interface forms between
the two phases.
Ammonia hydrolyzes the silicone oil, permitting a small concentration
to dissolve in the aqueous phase.
The sample is sealed and then 
mounted on the stage of a custom-built
holographic microscope with integrated holographic trapping
capabilities.

Imaging is performed with a collimated laser beam at
a vacuum wavelength of \SI{447}{\nm} (Coherent Cube).
Light scattered by the spheres interferes with the rest of
the beam in the focal plane of an objective lens
(Nikon, $100\times$, numerical aperture 1.45, oil immersion)
that collects the light and relays the magnified interference
pattern to a video camera (NEC, TI-324A) that records its intensity
at \SI{29.97}{frames\per\second}.
The same objective lens is used to project optical traps into the
sample.  These traps operate at a vacuum wavelength
of \SI{1064}{\nm} (IPG Photonics YLR-10-1064-LP) and are formed with
computer-generated holograms that are projected with
a phase-only spatial light modulator (Holoeye Pluto).

Holographic snapshots of colloidal particles are analyzed\cite{lee07a} using predictions
of the Lorenz-Mie theory of light
scattering\cite{bohren83,*mishchenko02} 
to obtain the spheres'
positions in three dimensions, their radii and their refractive
indexes.
Experiments on similar systems confirm nanometer precision
for tracking in-plane, 5 nanometers precision along the axial
direction, 3 nanometers precision for the spheres' radii, and
part-per-thousand resolution for their refractive indexes%
\cite{lee07a,cheong09,shpaisman12,krishnatreya14}.
Fit values for the refractive index are useful for distinguishing
particles on the basis of their composition\cite{lee07a,yevick14}.
Each hologram can be analyzed during the interval
between camera exposures\cite{yevick14}.

PDMS probe particles and silica spheres are dispersed in the aqueous phase 
at a volume fraction of 
\num{e-5} so that no more than one sphere appears
in the microscope's $\SI{86}{\um} \times\ \SI{65}{\um}$ field of view
at a time.
One PDMS sphere and one silica sphere are located, trapped in 
optical tweezers, and moved
under software control to the midplane of the channel.
The pair of spheres then is translated to a distance $x$ from
the oil-water interface.

\begin{figure}[!t]
  \centering
  \includegraphics[width=0.75\columnwidth]{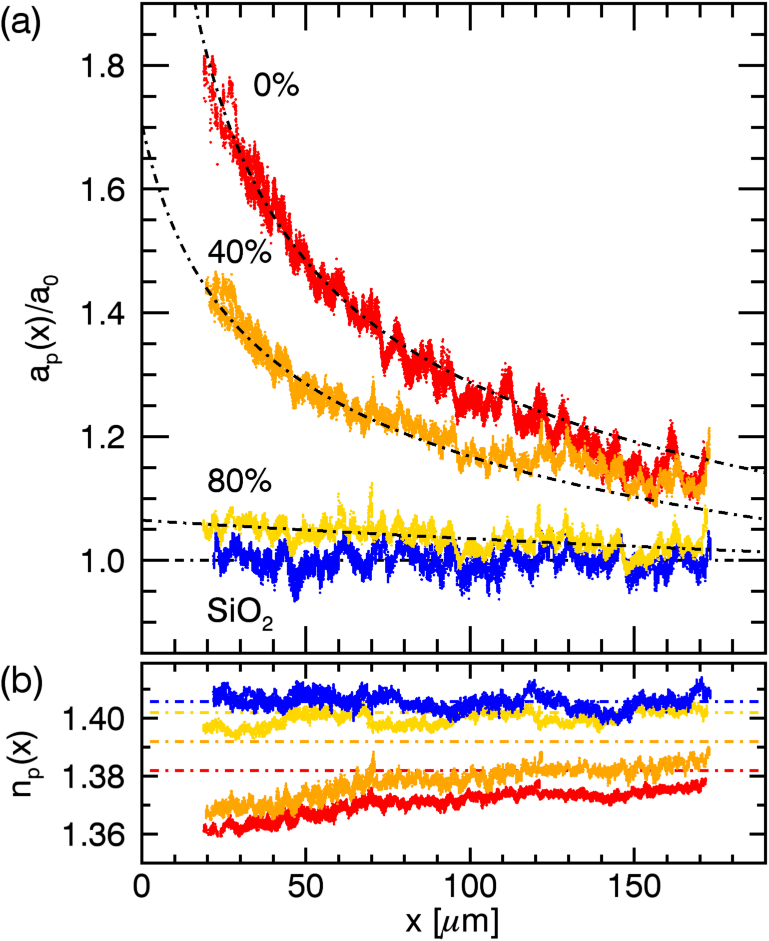}
  \caption{Holographically measured (a) radius $a_p$ and (b) refractive index $n_p$
    as a function of separation $x$ from the oil-water interface.
    Data are presented for $\xi = \num{0}$, \num{0.4} and \num{0.8},
    and for a \SI{2}{\um}-diameter silica sphere measured
    in tandem with the \SI{0}{\percent} PDMS sphere.  Overlaid
    curves in (a) are fits to Eq.~\eqref{eq:characteristic}.
    Horizontal lines in (b) show measured values of $n_0$.}
  \label{fig:data}
\end{figure}

From this point, we probe the concentration profile of monomers in
solution by moving the particles
back and forth relative to the interface, recording and analyzing 
their holograms as they move.
A detail from a typical holographic snapshot
is reproduced in
Fig.~\ref{fig:schematic}.
The corresponding Lorenz-Mie result for this hologram
illustrates the quality of a typical fit.
The data in Fig.~\ref{fig:data} show typical results 
from PDMS spheres with no
crosslinker (\SI{0}{\percent} MTES), \SI{40}{\percent} crosslinker
and \SI{80}{\percent} crosslinker, together with
control data for
a silica sphere
(SiO$_2$), that were obtained along with the \SI{0}{\percent} data.
Each trace in Fig.~\ref{fig:data} consists of more than 10,000
individual measurements of the particle's radius $a_p$ and refractive
index $n_p$
obtained in at least one complete cycle of moving back and forth between
$x = \SI{20}{\um}$ and $x = \SI{170}{\um}$.

Because silica is hydrophilic, silicone monomers and
oligomers should not wet the silica sphere's surface.  
A silica sphere's radius and refractive index
consequently should not vary with
monomer concentration.
Indeed, Fig.~\ref{fig:data} shows that the
radius and refractive index of the silica sphere
remains constant at $a_s =\SI{0.98(2)}{\um}$ and
$n_s = \num{1.406(3)}$, respectively,
independent of distance from the interface.
Comparable results were obtained for silica spheres moving
along with the crosslinked PDMS probe particles.
These control measurements
confirm that position-dependent variations in
the refractive index of the medium are too small to
influence holographic characterization in this system%
\cite{lee07a,moyses13,shpaisman12}.
This, in turn, confirms that position-dependent changes in
the observed properties of the PDMS probe particles reflect
changes in the particles themselves, and not artifacts due
to spatial variations in imaging conditions.

The PDMS spheres all swell consistently, reversibly and
reproducibly when passed back and forth through the
concentration gradient at a steady translation speed
of $v = \SI{0.7}{\um\per\second}$.
Their radii increase as they approach the
interface, and decrease as they move away.
Whereas the most highly crosslinked sphere swells by only
a few percent, the sphere with no crosslinker nearly doubles
its radius at its point of closest approach.
The results' repeatability suggests that concentration
gradients evolves very little over the course of the measurement.

\begin{figure}[!t]
  \centering
  \includegraphics[width=0.8\columnwidth]{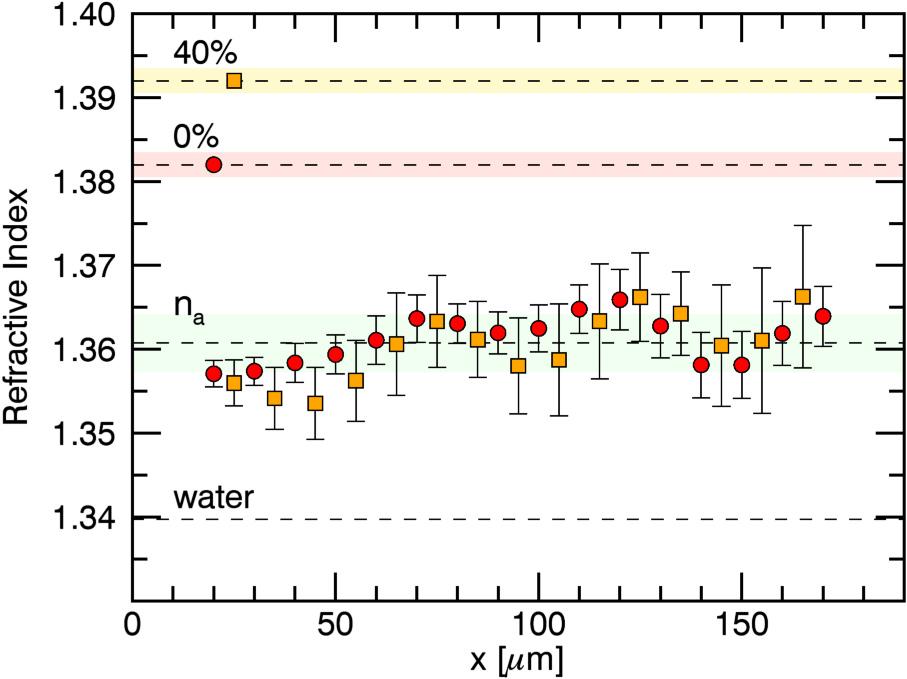}
  \caption{Absorbate refractive index, $n_a$ estimated with
    Eq.~\eqref{eq:lorentzlorenz}
    from the data in Fig.~\protect\ref{fig:data} for probe spheres with
   \SI{0}{\percent} and \SI{40}{\percent} crosslinker.
   Horizontal dashed lines indicate the refractive index of bulk PDMS
   with \SI{0}{\percent} and \SI{40}{\percent} crosslinker, and the
   refractive index of water.}
  \label{fig:refractiveindex}
\end{figure}
As the PDMS spheres swell, their refractive indexes decrease
slightly.  This would be explained naturally if the hydrolyzed
absorbate has a
lower refractive index than the fully dense gel matrix and
thus reduces the swollen spheres' mean refractive indexes\cite{shpaisman12}.
The influence of an absorbate of refractive index $n_a$ on the
overall refractive index $n_p$ of a swollen particle may be estimated
with effective medium theory\cite{aspnes82,*born99}:
\begin{equation}
  \label{eq:effectivemedium}
  f(n_p) = \phi_p \, f(n_0) + (1 - \phi_p) \, f(n_a),
\end{equation}
where $n_0$ is the refractive index of the unswollen sphere,
$\phi_p = (a_0/a_p)^3$ is the volume fraction of polymer within a
swollen sphere of radius $a_p$ whose unswollen radius is $a_0$,
and where
$f(n) = (n^2 - 1)/(n^2 + 2)$ is the Lorentz-Lorenz factor for a medium
of refractive index $n$.
The Lorentz-Lorenz factor for the absorbate may be estimated from the
data in Fig.~\ref{fig:data} as
\begin{equation}
  \label{eq:lorentzlorenz}
  f(n_a) = \frac{f(n_p(x)) \, a_p^3(x) - f(n_0) \, a_0^3}{a_p^3(x) - a_0^3},
\end{equation}
and, from this, the refractive index of the absorbate.
We measured $a_0$ and $n_0$ for each of the PDMS spheres by
translating them to $x = \SI{1}{\mm}$.

The data in Fig.~\ref{fig:refractiveindex} show results of this
analysis for the probe particles with \SI{0}{\percent} and \SI{40}{\percent}
crosslinker from Fig.~\ref{fig:data}.
The estimated absorbate refractive index is substantially independent
of position in the sample cell, as expected, despite large variations
in the particles' volumes over this range.
Results consistent with $n_a = \num{1.361(3)}$ are obtained from
both probe particles, despite their substantially different bulk
refractive indexes.
The success of this procedure provides additional support
for the accuracy and precision of the underlying holographic
characterization measurements.
The extracted value for the absorbates' refractive index falls
between the value for water, \num{1.340}, and that
for pure DMDES, \num{1.381}.

To verify that the optical trap itself does not influence
the sphere's properties, we moved the particles to
$x = \SI{20}{\um}$
from the interface, and extinguished the laser traps
to allow the particles
to diffuse freely.
Under these conditions, each probe particle's radius
increases and decreases in accord with its distance from the oil-water
interface, and in agreement with the data in Fig.~\ref{fig:data}.

The implicit relationship between the probe spheres' radii
and the local concentration of dissolved monomer can be made quantitative using 
the Flory-Rehner model for polymer swelling\cite{flory43a,*flory50,quesadaperez11}.
To do so, we treat a polymer sphere as consisting of a fixed
number, $N_p$, of bound monomers, each of specific volume $v_m$.
Swelling the sphere with $N_m$ free monomers increases its radius to
\begin{equation}
  \label{eq:ap}
  a_p = \left[\frac{3v_m(N_p + N_m)}{4\pi} \right]^{\frac{1}{3}}.
\end{equation}
The unswollen radius, $a_0$, corresponds to $N_m = 0$.

The tendency of free monomers to
be absorbed by the polymer matrix may be accounted for, in part,
by the entropy of mixing\cite{flory41,*huggins41,quesadaperez11}.
In addition to free and bound monomers, we assume that a fraction
$\xi$ of the bound monomers are trivalent crosslinkers.
Treating these three classes of monomers
as ideal gas molecules, their Gibbs free energy is
\begin{multline}
  \label{eq:Gentropy}
  \beta G_\text{FH}
  = 
  N_m \ln \phi_m 
  + \frac{1 - \xi}{\nu} \, N_p \ln\!\left((1 - \xi) \phi_p\right) \\
  + \frac{2}{3} \frac{\xi}{\nu} \, N_p \ln\!\left(\xi \phi_p\right),
\end{multline}
where 
$\beta^{-1} = k_B T$ is the thermal energy scale at absolute
temperature $T$ and where
$\phi_m = 1 - \phi_p$ is the volume fraction of free monomers in the sphere.
The factor of $\nu$ accounts for the loss of entropy of the bound
monomers due to their inability to rearrange freely.
The associated entropic contribution to the free monomers' chemical potential
in a sphere of radius $a_p$ is
\begin{align}
  \label{eq:muentropy}
  \beta \mu_\text{FH}(a_p) 
  & = 
    \beta \frac{dG_\text{FH}}{d N_m} \nonumber \\
  & = 
  \left(
    \frac{\nu - 1}{\nu}
    +
    \frac{1}{3} \frac{\xi}{\nu} 
  \right) \frac{a_0^3}{a_p^3} 
  + 
  \ln\!\left(1 - \frac{a_0^3}{a_p^3}\right).
\end{align}
Equations~\eqref{eq:Gentropy} and \eqref{eq:muentropy} do not account for the
energy of association between the free and bound monomers
that typically appears in Flory-Huggins theory%
\cite{flory41,*huggins41,quesadaperez11}.
This contribution may be ignored because it should be comparable to
the energy
of association between free monomers themselves,
and so should not
influence the exchange of monomers between the sphere and a reservoir
of monomers.

Swelling reduces the crosslinkers' entropy
and thereby contributes
\begin{equation}
  \label{eq:Gelastic}
  \beta G_e = \frac{3}{2} \, \frac{\xi}{\nu} \, N_p \left(\phi_p^{-\frac{2}{3}} - 1 \right)
\end{equation}
to the sphere's elastic free energy\cite{flory43a,*flory50,quesadaperez11}.
This in turn increases the chemical potential of absorbed monomers by
\begin{equation}
  \label{eq:muelastic}
  \beta \mu_e(a_p)
  = 
  \frac{\xi}{\nu} \, \frac{a_0}{a_p}.
\end{equation}

Finally, the interface between the sphere and the surrounding solution
has a surface tension $\gamma$ that raises the monomers'
free energy by
\begin{equation}
  \label{eq:Ggamma}
  G_s
  =
  4 \pi a_p^2 \gamma.
\end{equation}
We assume that the same value of $\gamma$ characterizes
the surface tension for free monomers, bound monomers and crosslinkers
alike.
Surface tension then contributes
\begin{equation}
  \label{eq:mugamma}
  \beta \mu_s(a_p) = \alpha \frac{a_0}{a_p}
\end{equation}
to the chemical potential, 
where $\alpha = 2 \beta v_m \gamma/a_0$ is the sphere's
thermal capillary number.

Assuming no other significant contributions, the chemical potential
of monomers in the sphere is
\begin{equation}
  \label{eq:chemicalpotential}
  \mu(a_p) = \mu_\text{FH}(a_p) + \mu_e(a_p) + \mu_s(a_p).
\end{equation}
This is related to the equilibrium concentration of free monomers at the sphere's
surface by
\begin{equation}
  \label{eq:concentration}
  c_s(a_p) = c_0 \, e^{\beta \mu(a_p)},
\end{equation}
where $c_0$ is the concentration at a planar interface.

When the sphere is placed in a solution with
concentration $c \leq c_0$, its radius is selected by the condition%
\cite{webster98}
\begin{equation}
  \label{eq:concentrationcondition}
  c_s(a_p) = c,
\end{equation}
and thus is a solution of the characteristic equation
\begin{equation}
  \label{eq:characteristic}
  \frac{c}{c_0}
  = 
  \left(1 - \frac{a_0^3}{a_p^3}\right)
  e^{
    \frac{\xi}{\nu} \left(\frac{a_0}{a_p} + \frac{1}{3} \frac{a_0^3}{a_p^3}\right)
    +
    \alpha \frac{a_0}{a_p} 
    +
    \frac{\nu - 1}{\nu}
    \frac{a_0^3}{a_p^3}
  }.
\end{equation}

The three adjustable parameters
in Eq.~\eqref{eq:characteristic},
can be assessed
independently.
The crosslinker fraction, $\xi$, is
determined during synthesis to within \SI{5}{\percent}.
The surface tension between silicone oil and the ammonia solution,
$\gamma = \SI{14(2)}{\milli\newton\per\meter}$,
is obtained with a pendant-drop tensiometer (attension, Theta Lite).
Given the monomers' specific volume of
$v_m \approx \SI{2.8e-28}{\cubic\meter}$, this yields a thermal
capillary number of $\alpha = \num{0.002}$, which is small
enough to neglect.
For the PDMS spheres in this study, independent NMR measurements%
\cite{obey94,*goller97}
suggest $\nu \approx 4$.
With these inputs, Eq.~\eqref{eq:characteristic} relates
a probe particles' radius
to the local monomer concentration $c(x)/c_0$ provided that the
particle remains in equilibrium with the solution.

\begin{figure}[!t]
  \centering
  \includegraphics[width=0.75\columnwidth]{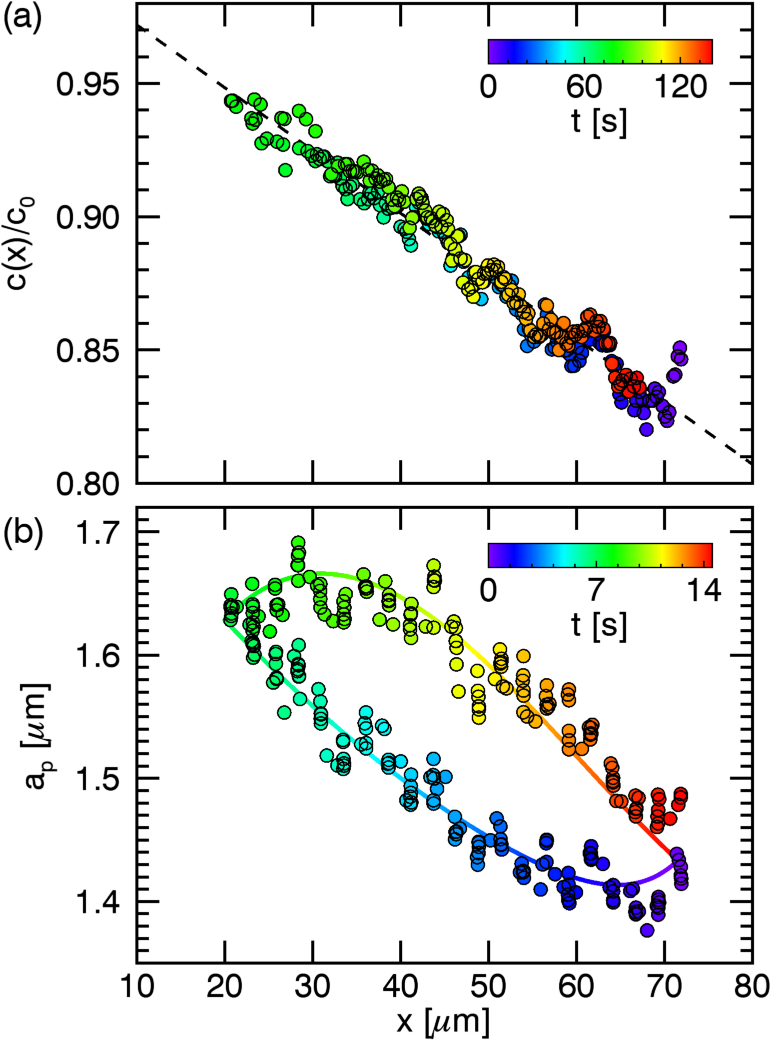}
  \caption{(a) Estimated concentration profile in a quasistationary
    concentration gradient measured with a probe particle moving
    at $v = \SI{0.7}{\um\per\second}$.  The results' reversibility
    confirms both that the sphere remains
    in equilibrium with the monomer bath and also that the
    concentration gradient does not change during the measurement.
    (b) Hysteresis in the response of the
    probe particle's size at $v = \SI{7}{\um\per\second}$. The continuous curve is
    computed with Eqs.~\eqref{eq:characteristic} and \eqref{eq:lsw}
    using the profile, $c(x)/c_0$, measured in (a).  Color
    corresponds to time.}
  \label{fig:gradientrate}
\end{figure}

Figure~\ref{fig:gradientrate}(a) shows a typical concentration
profile measured in this way using
a PDMS probe particle with \SI{0}{\percent}
crosslinker moving at $v = \SI{0.7}{\um\per\second}$.
The gradient is linear 
and passes through $c(0)/c_0 = 1$,
as expected.
Because the data in Fig.~\ref{fig:data} were obtained under similar
conditions, we can compare measured probe-particle radii with
predictions of Flory-Rehner theory
by numerically inverting Eq.~\eqref{eq:characteristic},
assuming a linear concentration profile.
Measurements of $a_p(x)$  plotted in
Fig.~\ref{fig:data}(a)
agree well with this model.
Estimates for $c(x)$ based on Eq.~\eqref{eq:characteristic}
should be less reliable for more highly crosslinked particles
both because of their more limited dynamic range, and also because
Eq.~\eqref{eq:characteristic} does not account for enthalpic
contributions to the elastic energy.

Having measured a concentration gradient, we then can probe the
behavior of the system when particles move too quickly to equilibrate
with the local concentration of monomers.
The data in Fig.~\ref{fig:gradientrate} were obtained with the probe
particle moving at a translation speed of $v =
\SI{7}{\um\per\second}$.
Rather than retracing the history of swelling and deswelling,
this trajectory displays substantial hysteresis.

We model this with the Lifshitz-Slyozov-Wagner equation%
\cite{lifshitz61,*wagner61,webster98},
\begin{equation}
  \label{eq:lsw}
  \frac{d a_p}{dt} = D v_m \, \frac{c(x(t)) - c_s(a_p(t))}{a_p(t)},
\end{equation}
where $D$ is the monomers' diffusion coefficient in solution.
For simplicity, we assume that $D$ is independent of concentration
over the relevant range of concentrations.
Using experimental data for $c(x)/c_0$ 
yields a prediction for the trajectory-dependent trace of $a_p(t)$ that is 
overlaid on the experimental data
in Fig.~\ref{fig:gradientrate}(b).
The sole adjustable parameter in this fit is the prefactor,
$c_0 v_m D$.

If the monomer diffusion coefficient is known independently
the analysis in Fig.~\ref{fig:gradientrate}
yields a measurement of $c_0$ and thus an absolute
measurement of $c(x)$.  
We estimate $D = \SI{e3}{\square\um\per\second}$ 
using the Stokes-Einstein relation
to obtain $c_0 = \SI{e7}{\per\cubic\um} = \SI{15}{\milli M}$.
This is below the current resolution limit\cite{shpaisman12}
for holographic measurement of dissolved species through their
influence on the medium's refractive index.
It is consistent therefore with the earlier observation
that estimates for particles' refractive indexes do not vary
appreciably along the concentration gradient in the sample
cell.
Stimulus-responsive probe particles therefore extend
the sensitivity of holographic microrefractometry\cite{shpaisman12}
for measuring small concentrations.

This study demonstrates that holographic video microscopy is
effective for measuring the response of stimulus-responsive
sensor particles.   Our particular implementation uses swelling of
polymer-gel particles to monitor the concentration of monomers 
in solution.  Probe particles with different functionality that swell in
response to other environmental factors could be used just
as easily.  Indeed, a variety of stimulus-responsive probes could
be deployed in a single system to monitor multiple
physical and chemical factors simultaneously.  Holographic characterization's ability
to distinguish particle types by size and refractive index would
enable parallel readout while maintaining speed and precision.

This work was supported by the National Science Foundation
through Grant Number DMR-1305875.  The holographic trapping
and characterization instrument was developed under the MRI program
of the NSF through Grant Number DMR-0922680.  Additional support
was provided by a grant from Procter \& Gamble.

\end{document}